\documentstyle[12pt,epsf]{article}

\begin{document}

\begin{titlepage}

\begin{flushright}
ICRR-Report-384-97-7 \\
KUNS-1437 HE(TH)97/04 \\
hep-ph/9703321
\end{flushright}

\begin{center}
\vspace*{1.2cm}

{\LARGE\bf Infra-Red Fixed Points \\[6mm] in an Asymptotically
  Non-Free Theory }
\vspace*{1.5cm}

{\large Masako {\sc Bando} \footnote{ e-mail address: {\tt
      bando@aichi-u.ac.jp}}}
\vspace{0.2cm}

{\it Aichi University, Miyoshi, Aichi 470-02, Japan}
\vspace{0.6cm}

{\large Joe {\sc Sato} \footnote{e-mail address: {\tt
      joe@hep-th.phys.s.u-tokyo.ac.jp}}}
\vspace{0.2cm}

{\it Institute for Cosmic Ray Research \\ The University of Tokyo,
Midori-Cho, Tanashi, Tokyo 188, Japan}
\vspace{0.6cm}

{\large Koichi {\sc Yoshioka} \footnote{e-mail address: {\tt
      yoshioka@gauge.scphys.kyoto-u.ac.jp}}}
\vspace{0.2cm}

{\it Department of Physics, Kyoto University \\ Kyoto 606-01, Japan }
\vspace*{1.5cm}

{\large\bf Abstract}
\begin{quote}
\hspace*{2ex} We investigate the infrared fixed point structure in
asymptotically free and asymptotically non-free theory. We find that
the ratios of couplings converge strongly to their infrared fixed
points in the asymptotically non-free theory.
\end{quote}
\end{center}
\end{titlepage}

\newpage

\section{Introduction}
\setcounter{equation}{0}\setcounter{footnote}{0}

\hspace*{2ex} The manner in which the Yukawa couplings reproduce the
masses of quarks and leptons is one of the unsolved problems in
particle physics. These masses may be constrained by requiring some
symmetry or by imposing grand unified theory and/or they may be
related to gauge couplings.  The latter is strongly indicated by the
top quark mass.
About 3 years ago, Lanzagorta and Ross \cite{Ross} attempted to
determine the relation between Yukawa couplings and gauge couplings
through the infrared fixed point structure. This method in which the
infrared fixed points may determine the heavy fermion masses was first
proposed by Pendleton and Ross \cite{pendletonross}. \ However, such
an infrared fixed point has rarely been reached in usual
asymptotically free (AF) standard models because of the infrared
divergent character \cite{hill}. \ Furthermore, since the coupling
constants become very large in the low energy region, the perturbative
treatment is no more guaranteed in the infrared region.

More than 40 years ago, Landau proposed an attractive idea that low
energy physics may be determined with an asymptotically non-free (ANF)
theory. He illustrated this idea by showing the possibility to obtain
a very small fine structure constant. This idea was applied to
determine the ratio of gauge couplings \cite{MPP}. \ In particular, it
was pointed out by Moroi {\it et al.}\ \cite{MMY} that the minimal
supersymmetric standard model (MSSM) with 1 extra vector-like
family (EVF) gives the observed ratio of the gauge couplings (Weinberg
angle).

In a previous paper \cite{Bando}, \ we investigated a possible
scenario of the standard gauge symmetry with ANF character and showed
that due to the ANF gauge couplings the top Yukawa coupling is quite
insensitive to their initial values fixed at GUT scale $M_G$. We would
like to stress that such strong convergence of Yukawa couplings to
their infrared fixed points is a common feature appearing in ANF
theories.

In this paper we investigate how strongly the couplings are focused
into their infrared points in ANF theories and demonstrate the
structure of the renormalization-group flow.  As illustrations we take
the supersymmetric standard models with AF and ANF gauge couplings and
compare them by concentrating on their infrared structure.

\vspace*{0.5cm}

\section{Infrared Structure of AF and ANF Theories}
\setcounter{equation}{0}\setcounter{footnote}{0}
\addtocounter{footnote}{1}

\begin{figure}[htbp]
  \begin{center}
    \epsfxsize=9cm \epsfysize=7.9cm \ \epsfbox{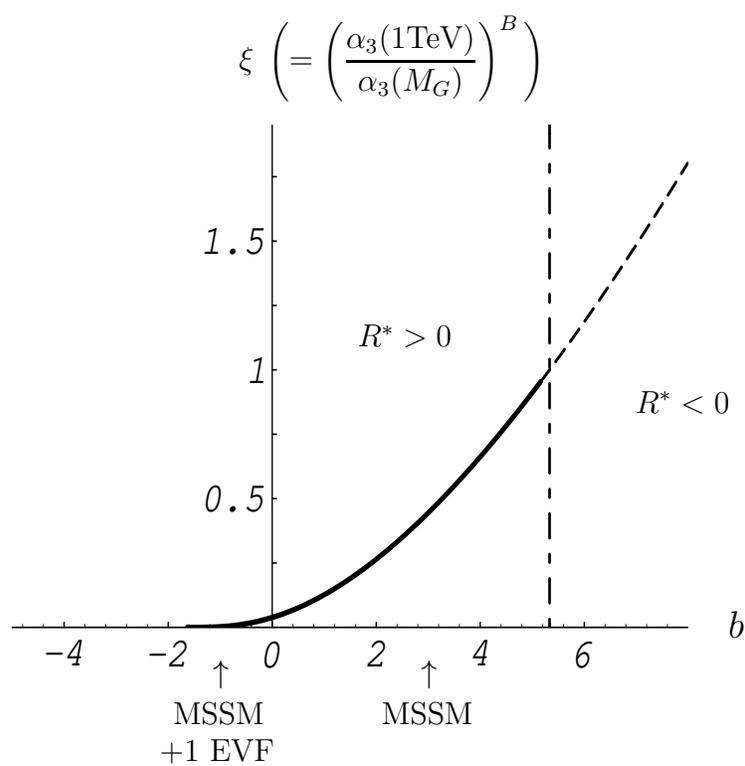}
    \put(-170,235){$\xi\;
      \left( = \Biggl
          ( \displaystyle{\frac{\alpha_3(1\mbox{TeV})}{\alpha_3(M_G)}}
        \right)^B \,\Biggr)$}
    \put(-20,105){$R^* < 0$}
    \put(-125,130){$R^* > 0$}
    \put(15,20){\large $b$}
    \put(-180,3){$\uparrow$}
    \put(-101,3){$\uparrow$}
    \put(-195,-13){MSSM}
    \put(-200,-27){$+1$ EVF}
    \put(-116,-13){MSSM}
    \caption{Typical behavior of the suppression factor $\xi$ by
      taking $\alpha = \alpha_3$ }
    \vspace*{2mm}
    \hspace*{-1cm}$( \alpha_3(M_Z) = 0.12\,,\,c = \frac{16}{3} )$
    \label{fig:sup}
  \end{center}
\end{figure}

\hspace*{2ex} Before studying realistic (AF and ANF) models we first
consider a simple gauged Yukawa system which has one gauge coupling
$g$ and one Yukawa coupling $y$, whose 1-loop $\beta$-functions are

\begin{eqnarray}
  \frac{d \alpha}{d t} &\!\! = \!\!& - \frac{b}{2\pi} \alpha^2,
  \\[2mm]
  \frac{d \alpha_y}{d t} &\!\! = \!\!& \frac{\alpha_y}{2\pi}\, \Bigl
  ( a \alpha_y - c \,\alpha \Bigr),
\end{eqnarray}
where
\begin{eqnarray}
  \alpha \equiv \frac{g^2}{4\pi} \,,&& \alpha_y \equiv
  \frac{y^2}{4\pi}\,,\qquad t = \ln \left(\frac{\mu}{\mu_0}\right) .
\end{eqnarray}
The system is asymptotically free (asymptotically non-free) for
$b > 0  \;( b < 0 ) $ and always $a > 0\,,\,c \geq 0 $.
{}~From these, we obtain

\begin{eqnarray}
  &&\frac{d R}{d t} = \frac{a}{2 \pi}\alpha R \left( R - R^*
  \right)\,, \qquad\;  \left( R \equiv \frac{\alpha_y}{\alpha} \right)
  \label{eqn:R}\\[2mm]
  &&\hspace*{2cm}R^* = \frac{c-b}{a}.
\end{eqnarray}
Since we would like to see the manner in which couplings reach
infrared fixed points from their values at the GUT scale $M_G\,$, we
take $\,\mu_0 = M_G$ and $\alpha(0) = \alpha(M_G)$. Then we obtain
from (\ref{eqn:R}) \cite{Bando}

\begin{eqnarray}
  \frac{R(t) - R^*}{R(t)} &\!\! = \!\!&
  \left(\frac{\alpha(t)}{\alpha(M_G)}\right)^B \left( \frac{R(M_G)
      - R^*}{R(M_G)}\right)\,,
      \label{eqn:Rsol}
\\[2mm] && \; B \equiv 1 - \frac{c}{b}.
\end{eqnarray}
Here, $R^*$ is an infrared fixed point if $R^*>0$. This is equivalent
to what Lanzagorta and Ross derived in Ref.~\cite{Ross}. \ From this
equation, we see that the suppression factor $\xi \equiv \left(
\frac{\alpha (t)}{\alpha(M_G)} \right)^B$ provides the criterion on
rate at which $R$ approaches the infrared fixed point value
$R^*$. Note that this factor is independent of the detailed
information of the system and is determined from gauge coupling
alone. The $b$-dependence of the suppression factor $\xi$ is shown in
Fig.~\ref{fig:sup},
$\,$from which we find a large difference between  AF ($b>0$) and ANF
($b<0$) cases. In the AF case the point $b=c$ at which $B$ becomes 0
corresponds to  $\xi=1$. If we dare to extrapolate our formula above
this critical point, $\xi$ becomes larger than 1, but $R^*$ becomes
negative and is no more an
infrared fixed point. From (\ref{eqn:Rsol})
we see that $\xi$ blows up and $R(t)$ tends to zero in the low energy
limit \cite{kugo}. \ On the other hand, in the ANF case there is
always a nontrivial infrared fixed point $R^* (>0)$, and convergence
to $R^*$ becomes much better.

\begin{figure}[htbp]
  \begin{center}
    \epsfxsize=9cm \epsfysize=7.9cm \ \epsfbox{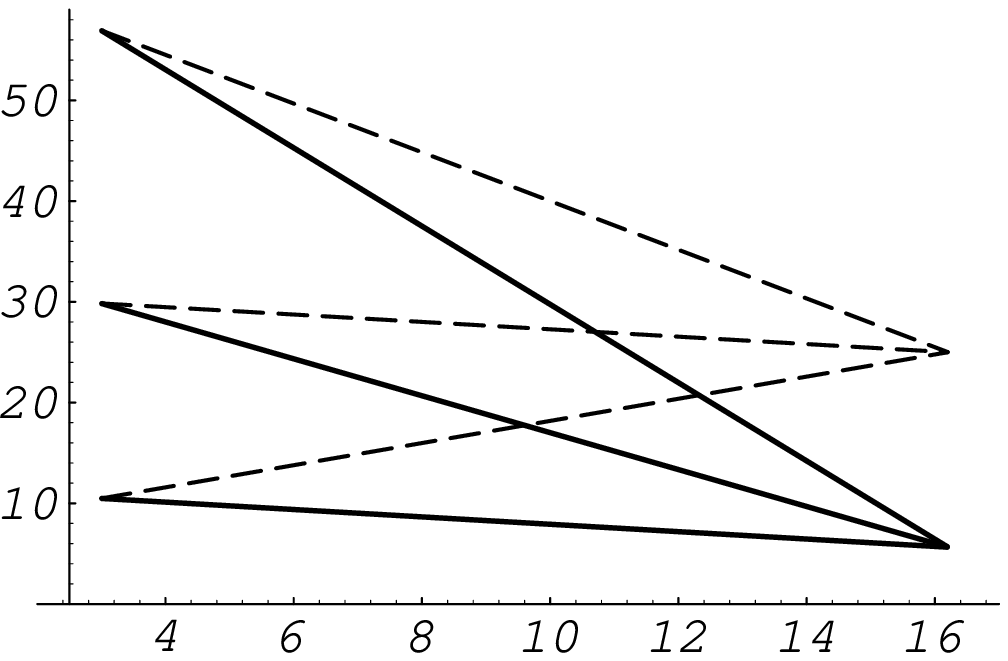}
    \put(-245,225){\large $\alpha^{-1}$}
    \put(15,25){ $\log_{10} \left(\mu/[\mbox{GeV}]\right)$}
    \caption{Typical $\mu$-dependence of
      $\alpha_1(\mu)\,,\,\alpha_2(\mu)\,,\,\alpha_3(\mu)$ ~in the MSSM
      (dashed lines) and the MSSM + 1 EVF (solid lines).}
    \vspace*{-0.5cm}
    \label{fig:alpha}
  \end{center}
\end{figure}

Now we illustrate the difference in behavior of the AF and ANF
theories by applying this formula to two models, the minimal
supersymmetric standard model and the MSSM with one extra vector-like
family (EVF) \cite{MPP,MMY,Bando}. \ Typical behavior of the running
gauge couplings is shown in Fig.~\ref{fig:alpha}.
They are unified at the same scale in the two cases, but with
different unified couplings.

Let us first exhibit typical values for the AF and ANF cases, taking
$\alpha = \alpha_3\,,\,\alpha_y = \alpha_t$ with a realistic value of
$\alpha_3$ (see \S 3 for details),

\begin{eqnarray}
  \mbox{MSSM} \hspace*{1cm}&:& ~~b = 3 \,,\quad c = \frac{16}{3}
  \quad\Rightarrow\quad  B = -\frac{7}{9} \nonumber\\[2mm]
  && \left( \frac{\alpha_3(M_Z)}{\alpha_3(M_G)} \right)^B \;\sim
  \;\left( \frac{0.12}{0.04} \right)^{-7/9}\; \sim\; 0.43\>, \\[2mm]
  \mbox{MSSM + 1 EVF~~} &:& ~~b = -1 \,,\quad c = \frac{16}{3} \quad
  \Rightarrow \quad B = \frac{19}{3} \nonumber\\[2mm]
  && \left( \frac{\alpha_3(M_Z)}{\alpha_3(M_G)} \right)^B \;\sim
  \;\left( \frac {0.12}{1.0} \right)^{19/3}\; \sim\; 10^{-6} \>.
\end{eqnarray}
The suppression factors can be read off Fig.~\ref{fig:sup}
(indicated by arrows). We can see the situation more clearly by
comparing the $\mu$-dependence of $\alpha_t/\alpha_3$ in the AF and
ANF cases. In the MSSM + 1 EVF, the convergence to the infrared fixed
point is much better than that in the MSSM, and its fixed point value
depends very weakly on the initial value at $M_G$ (Fig.~4).
This is because gauge couplings are asymptotically non-free and
their unified coupling is very large at $M_G$ (Fig.~2).

\begin{figure}[htbp]
  \begin{center}
    \epsfxsize=9cm \epsfysize=7.9cm \ \epsfbox{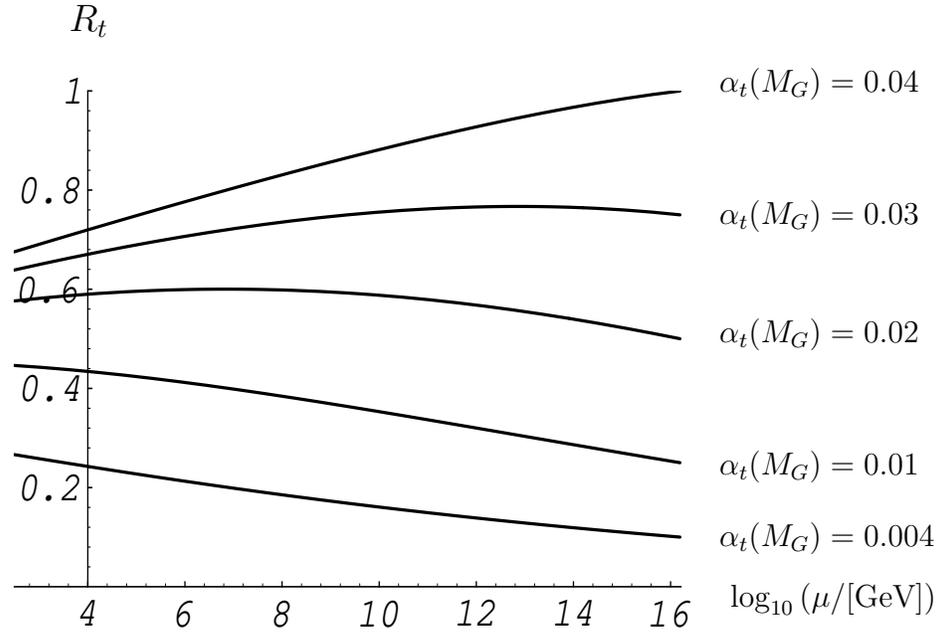}
    \put(15,15){{$\log_{10} \left(\mu/[\mbox{GeV}]\right)$}}
    \put(-235,233){{\large $R_t$}}
    \put(11,210){$\alpha_t (M_G) = 0.04$}
    \put(11,160){$\alpha_t (M_G) = 0.03$}
    \put(11,115){$\alpha_t (M_G) = 0.02$}
    \put(11,65){$\alpha_t (M_G) = 0.01$}
    \put(11,38){$\alpha_t (M_G) = 0.004$}
    \caption{$R_t$ ~in the MSSM ~
      ($M_G = 1.6\times 10^{16}\;\mbox{GeV}\,,\; \alpha_{GUT} =
      0.04$)}
  \end{center}
\end{figure}

\begin{figure}[htbp]
  \begin{center}
    \epsfxsize=9cm \epsfysize=7.9cm \ \epsfbox{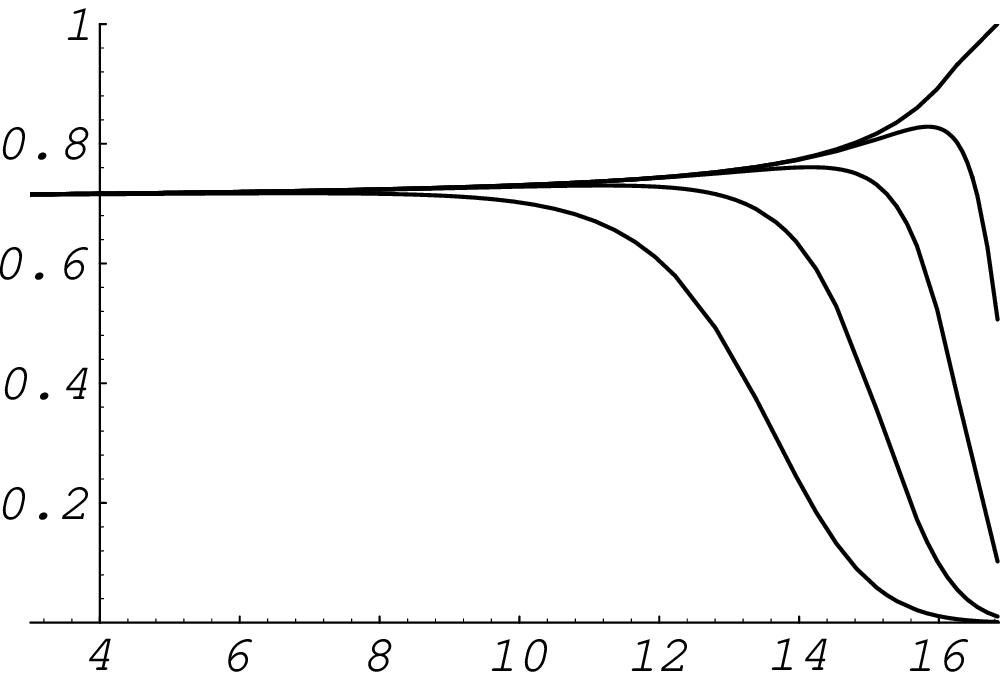}
    \put(-235,235){{\large $R_t$}}
    \put(11,210){$\alpha_t (M_G) = 1.0$}
    \put(11,115){$\alpha_t (M_G) = 0.5$}
    \put(11,38){$\alpha_t (M_G) = 0.1$}
    \put(11,23){$\alpha_t (M_G) = 0.01$}
    \put(11,8){$\alpha_t (M_G) = 0.001$}
    \caption{$R_t$ ~in the MSSM
      + 1 EVF ~($M_G = 7.0\times 10^{16}\;\mbox{GeV}\,,\; \alpha_{GUT}
      = 1.0$)}
    \label{fig:t/3-ANF}
  \end{center}
\end{figure}

\begin{figure}[htbp]
\vspace*{-5mm}
  \begin{center}
    \epsfxsize=9cm \epsfysize=8cm \ \epsfbox{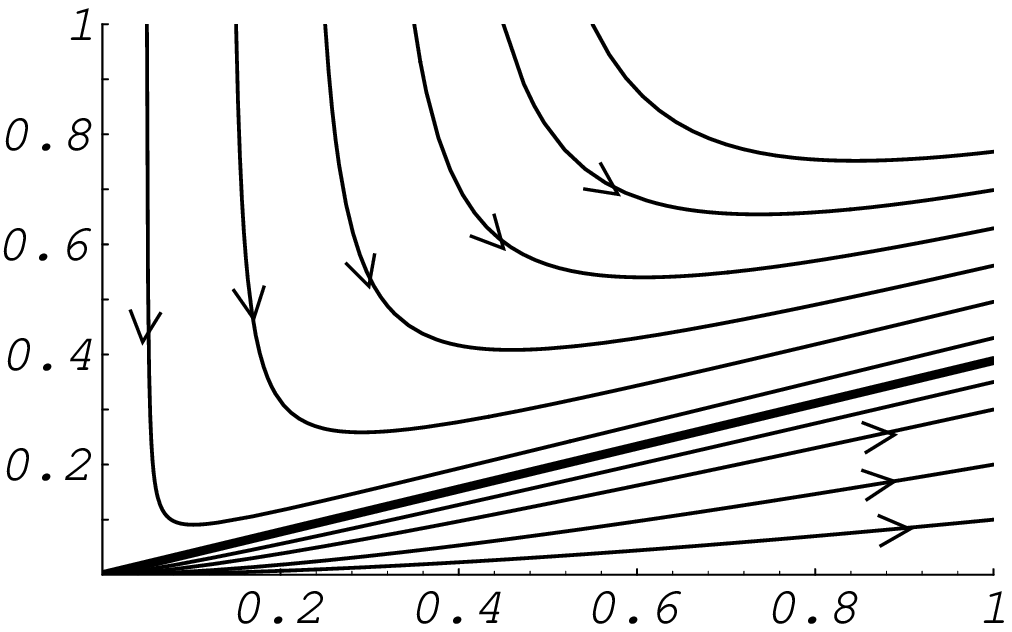}
    \put(12,24){{\large $\alpha_3$}}
    \put(-235,230){{\large $\alpha_t$}}
    \put(10,97){fixed line}
    \caption{RG flow diagram in the MSSM }
    \vspace*{0.2cm}~~~~(The arrows denote the flow directions toward
    the infrared region.)
    \label{fig:flowMSSM}
  \end{center}
\end{figure}

\begin{figure}[htbp]
\vspace*{-5mm}
  \begin{center}
    \epsfxsize=9cm \epsfysize=8cm \ \epsfbox{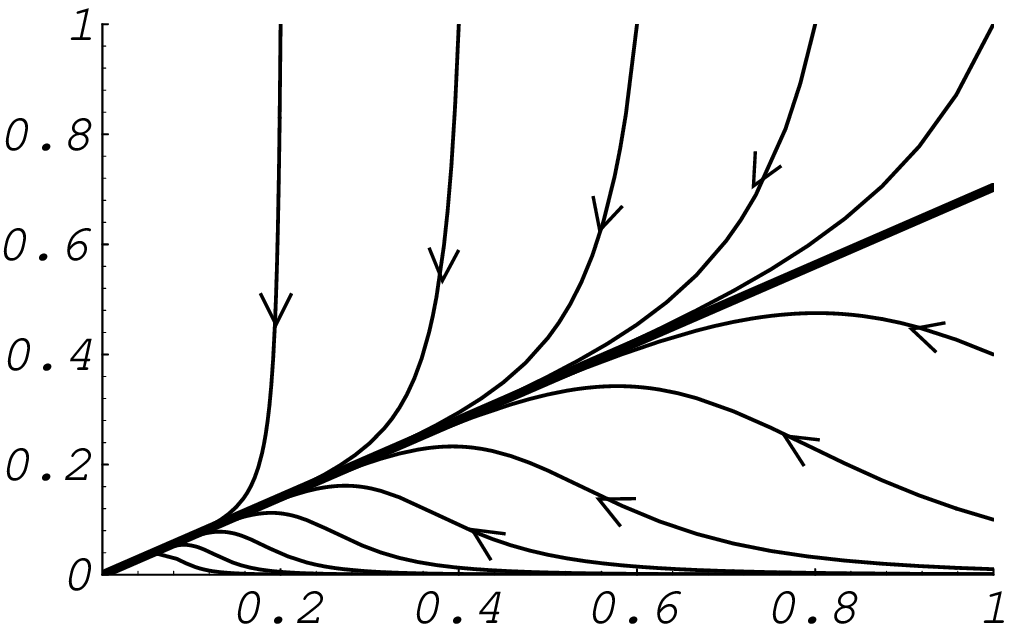}
    \put(15,22){{\large $\alpha_3$}}
    \put(-235,230){{\large $\alpha_t$}}
    \put(10,160){fixed line}
    \caption{RG flow diagram in the MSSM + 1 EVF }
    \vspace*{0.2cm}~~~~(The arrows denote the flow directions toward
    the infrared region.)
    \label{fig:flowANF}
  \end{center}
\end{figure}

Next we investigate the RG flow in the $(\alpha_3,\alpha_t)$ plane.
The RG flow behaves quite differently according to the regions
separated by the fixed line $\alpha_t/\alpha_3 = R_t^* $
(Figs.~5 and 6).

In the lower region of the AF case, the Yukawa coupling, as well as
the gauge coupling, is found to be asymptotically free, and we have a
nontrivial continuum limit \cite{kugo}. \ Although the ratio finally
approaches the infrared fixed point, in the infrared region both of
the couplings become very large, and the one-loop approximation is no
longer reliable.

For the other situations, either or both couplings are divergent at
high energy. In the lower region of the ANF case the gauge coupling
governs the Yukawa coupling very strongly. The gauge coupling diverges
at some high energy scale. In the upper region for both cases, on the
other hand $\alpha_t$ diverges at some high energy scale. Then the
theory becomes trivial in the continuum limit and we should make use
of cut-off theory.  The former corresponds to dynamical gauge
boson \cite{dgb} and the latter, dynamical Higgs
boson \cite{dynHiggs}.

It is noted that the ANF case is in strong contrast with the AF
case.  In both the upper and lower regions, the ratio evidently has an
infrared fixed point where the one loop approximation becomes more
and more valid. This can be clearly seen in Fig.~6.  This infrared
stability is very attractive and may make the determination of Yukawa
coupling constants feasible.

\vspace*{0.5cm}

\section{Infrared Solutions of MSSM + 1 EVF}
\setcounter{equation}{0}\setcounter{footnote}{0}

\hspace*{2ex} Keeping the above attractive features in mind, we
further undertake an analysis of the infrared fixed point solutions of
MSSM + 1 EVF, which includes three gauge and Yukawa couplings. We
suppose that the superpotential takes the form
\begin{eqnarray}
  W &=& W_3 + W_4 + W_{\bar{4}}\,, \\[2mm]
  &&\hspace*{-5mm}W_3 = y_{t_3} Q_3 H \bar{t}_3 + y_{b_3} Q_3 \bar{H}
  \bar{b}_3 + y_{\tau_3} L_3 \bar{H} \bar{\tau}_3\,, \\[2mm]
  &&\hspace*{-5mm}W_4 = y_{t_4} Q_4 H \bar{t}_4 + y_{b_4} Q_4 \bar{H}
  \bar{b}_4 + y_{\tau_4} L_4 \bar{H} \bar{\tau}_4\,, \\[2mm]
  &&\hspace*{-5mm}W_{\bar{4}} = y_{\bar{t}} \bar{Q} \bar{H}
  t_{\bar{4}} + y_{\bar{b}} \bar{Q} H b_{\bar{4}} + y_{\bar{\tau}}
  \bar{L} H \tau_{\bar{4}}\,,
\end{eqnarray}
and the Yukawa couplings satisfy \footnote{The relations are always
satisfied once we put these relations at $M_G$.}
\begin{eqnarray}
  y_{t_3} = y_{t_4} \equiv y_t \,,&&\quad y_{b_3} = y_{b_4} \equiv y_b
  \,,\quad \; y_{\tau_3} = y_{\tau_4} \equiv y_\tau \,,\label{eqn:3=4}
  \\[2mm]
  &&~y_{\bar{t}} = y_{\bar{b}} = y_{\bar{\tau}} = 0\,.
\end{eqnarray}
With these assumptions, there exist parameter regions where both
the low-energy experimental values and the high energy GUT-like
boundary conditions are consistently described \cite{Bando}. \ Then
the 1-loop $\beta$-functions for the ratio of the couplings to
$\alpha_3\,\left( R_i \equiv \alpha_i/\alpha_3\right)$ are

\begin{eqnarray}
  \frac{d R_1}{d t } &=& \frac{1}{2\pi} \alpha_3
  R_1 \bigl( 10.6 R_1 - 1 \bigr) \,,\label{eqn:alp1} \\[2mm]
  \frac{d R_2}{d t } &=& \frac{1}{2\pi} \alpha_3 R_2 \bigl(
  5 R_2 - 1 \bigr) \,,\label{eqn:alp2} \\[2mm]
  \frac{d R_t}{d t } &=& \frac{1}{2\pi} \alpha_3 R_t \left(
    9 R_t + R_b - \frac{13}{15}R_1 - 3 R_2 - \frac{19}{3} \right) \,,
  \label{eqn:alp3} \\[2mm]
  \frac{d R_b}{d t } &=& \frac{1}{2\pi} \alpha_3 R_b \left(
    R_t + 9 R_b + 2 R_\tau - \frac{7}{15}R_1 - 3 R_2 - \frac{19}{3}
   \right) \,,\label{eqn:alp4} \\[2mm]
   \frac{d R_\tau}{d t } &=& \frac{1}{2\pi} \alpha_3 R_\tau
   \left( 6 R_b + 5 R_\tau - \frac{9}{5}R_1 - 3 R_2\right)\,.
   \label{eqn:a lp5}
\end{eqnarray}

\begin{figure}[htbp]
  \begin{center}
    \epsfxsize=10cm \epsfysize=7.5cm \ \epsfbox{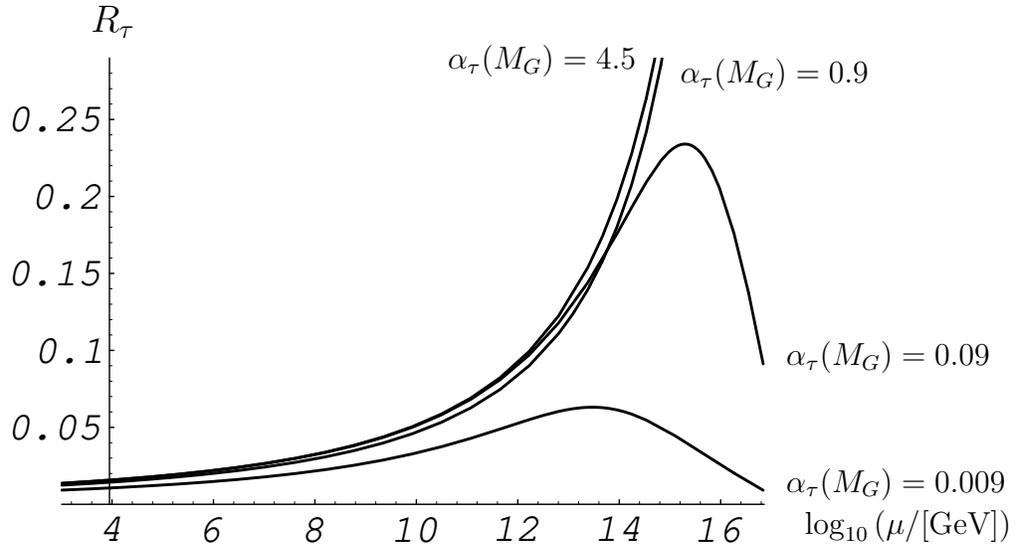}
    \put(15,17){$\log_{10} \left(\mu/[\mbox{GeV}]\right)$}
    \put(-255,207){\large $R_\tau$}
    \put(-120,193){$\alpha_\tau (M_G) = 4.5$}
    \put(-32,188){$\alpha_\tau (M_G) = 0.9$}
    \put(8,81){$\alpha_\tau (M_G) = 0.09$}
    \put(8,33){$\alpha_\tau (M_G) = 0.009$}
    \caption{ ~$R_\tau$ in the MSSM + 1 EVF ~ ($M_G = 7.0\times
      10^{16}\;\mbox{GeV}\,,\;\, \alpha_{GUT} = 1.0$)}
    \label{fig:tau}
  \end{center}
\end{figure}

First we comment with regard to the infrared structure of
$R_\tau$. The naive infrared fixed point solution from all these
$\beta$-functions should not be interpreted as the true fixed point,
since this solution of $R_\tau$ is negative ($R_\tau^*
\,\sim\,-0.66$). In this case, as was mentioned above, $R_\tau$ does
not reach the non-trivial infrared fixed point, but rather zero, in
the low energy limit (Fig.~\ref{fig:tau}).

In order to obtain the correct infrared fixed point solutions, we
first set $R_\tau \sim 0$ and use the four remaining $\beta$-functions
[(\ref{eqn:alp1}) $\sim$ (\ref{eqn:alp4})]. The results are as
follows.\footnote{Strictly speaking, $R_1$ and $R_2$
  do not converge to their fixed point values. However, the
  differences $R_i(1 \mbox{TeV}) - R^*_i \; (i= 1,2)$ are small enough
  for us to use the result (\ref{eqn:fpsol}).}

\begin{eqnarray}
  R_1^* = 0.0943\,,\qquad R_2^* = 0.2\,,
\end{eqnarray}
\vspace*{-0.9cm}

\begin{eqnarray}
  R_t^* = 0.702 \,,\qquad R_b^* =
  0.697\,. \qquad ( R_\tau \sim 0.0 )
\label{eqn:fpsol}
\end{eqnarray}

\begin{figure}[htbp]
  \begin{center}
    \epsfxsize=11cm \epsfysize=8cm \ \epsfbox{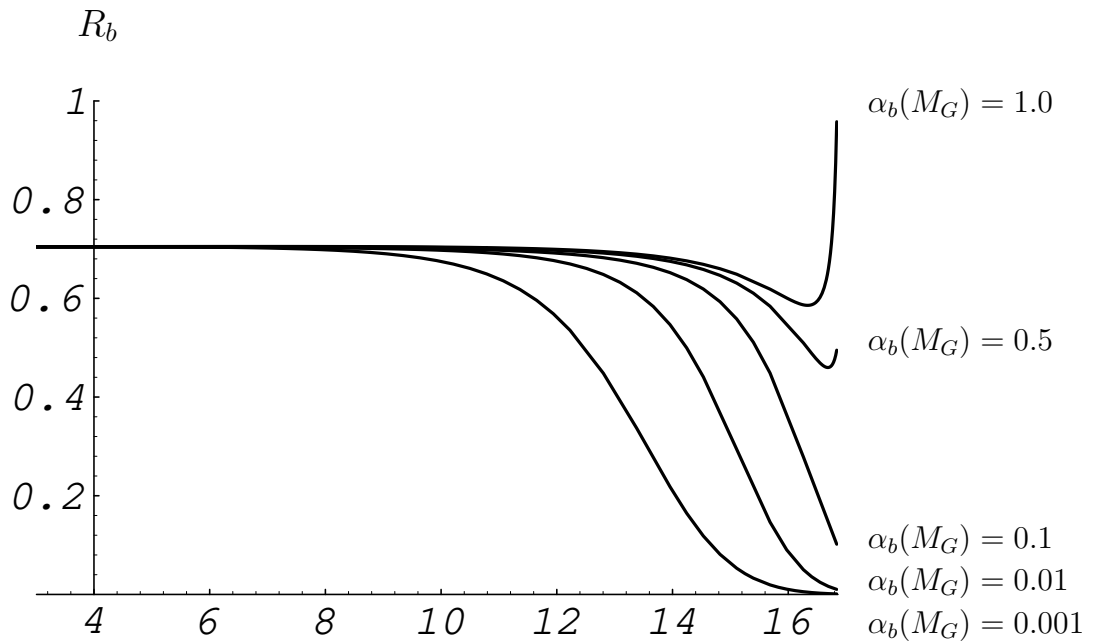}
    \put(-288,236){{\large $R_b$}}
    \put(11,207){$\alpha_b (M_G) = 1.0$}
    \put(11,117){$\alpha_b (M_G) = 0.5$}
    \put(11,42){$\alpha_b (M_G) = 0.1$}
    \put(11,26){$\alpha_b (M_G) = 0.01$}
    \put(11,10){$\alpha_b (M_G) = 0.001$}
    \caption{ ~$R_b$ in the MSSM + 1 EVF  ~($M_G = 7.0\times
      10^{16}\;\mbox{GeV}\,,\;\, \alpha_{GUT} = 1.0$)}
    \label{fig:bottom}
  \end{center}
\end{figure}

{}From Figs.~4 and \ref{fig:bottom}, \ we find that
these quantities indeed reach their fixed points. As we mentioned
before, these values are affected little by the initial values at
$M_G$ because of the asymptotically non-free gauge couplings.  We are
confident that the infrared fixed points obtained from these solutions
are physically significant and provide us with reliable low-energy
parameters.

In the case at hand, by using these fixed point solutions and the
experimental value of $\alpha_3(1 \mbox{TeV}) \sim \, 0.093$ ,
we obtain, for example,\footnote{The value $\tan \beta \sim 58$ is
determined from the experimental value $m_\tau(M_Z)$ \cite{Bando}. }
\begin{eqnarray}
  m_t(M_Z) \sim 178 \;\mbox{GeV}\,,&&\quad\; m_b(M_Z) \sim 3.2
  \;\mbox{GeV} \,.\\[1mm] &&\hspace*{-1cm}( \,\tan \beta \sim 58\, )
  \nonumber
\end{eqnarray}

\noindent These values are certainly consistent with the experimental
values \cite{exp}
\begin{eqnarray}
  m_t(M_Z) \sim 180 \pm 10 \;\mbox{GeV}\,,\quad m_b(M_Z) \sim 3.1 \pm
  0.4 \;\mbox{GeV}\,.
\end{eqnarray}

\vspace*{1cm}

\section{Conclusion}
\setcounter{equation}{0}\setcounter{footnote}{0}

We found interesting infrared structure which is commonly seen in ANF
theories. This is an important difference between AF and ANF
theories. This possibility has long been observed but has never
been taken serious.

The existence of extra fermions has been discussed from various points
of view: in deriving $CP$ violation, dynamical SUSY breaking,
hierarchical mass matrix, and so on. In particular, GUT models beyond
standard models, including string models and supergravity models,
predict additional fermions quite naturally. We can expect that at low
energy, the theory is asymptotically non-free.

We would like to stress that since ANF theories have strong predictive
powers because of the strong convergence of couplings (ratio of
couplings) to infrared fixed points, their study is considered
important.

\vspace*{0.5cm}

\section*{Acknowledgments}
The authors would like to express their sincere thanks to T.\ Kugo and
T.\ Yanagida for their valuable comments and encouragement. The
previous work with T.\ Takeuchi and T.\ Onogi (as well as J.\ S. and
M.\ B.) has provided a strong motivation for this work, and
discussions with them were instructive. A part of this work was
performed during the Kashikojima Workshop held in August 1996 at
Kashikojima Center, which was supported by a Grand-in Aid for
Scientific Research (No. 07304029) from the Ministry of Education,
Science and Culture. M.\ B. is supported in part by the Grand-in Aid
for Scientific Research (No.33901-06640416) from the Ministry of
Education, Science and Culture.

\end{document}